\documentclass[aps,prb,twocolumn,superscriptaddress,showpacs,showkeys,floatfix]{revtex4}

\usepackage{graphicx}
\graphicspath{{figs/}}
\begin{document}
\title{Superior thermal conductivity and extremely high mechanical strength in polyethylene chains from {\it ab initio} calculation}
\author{Jin-Wu~Jiang}
    \affiliation{Institute of Structural Mechanics, Bauhaus-University Weimar, Marienstr. 15, D-99423 Weimar, Germany}
\author{Junhua~Zhao}
    \affiliation{Institute of Structural Mechanics, Bauhaus-University Weimar, Marienstr. 15, D-99423 Weimar, Germany}
\author{Kun~Zhou}
    \affiliation{School of Mechanical and Aerospace Engineering, Nanyang Technological University, 50 Nanyang Avenue, Singapore 639798, Singapore}
\author{Timon~Rabczuk}
    \affiliation{Institute of Structural Mechanics, Bauhaus-University Weimar, Marienstr. 15, D-99423 Weimar, Germany}

\date{\today}
\begin{abstract}
The upper limit of the thermal conductivity and the mechanical strength are predicted for the polyethylene chain, by performing the {\it ab initio} calculation and applying the quantum mechanical non-equilibrium Green's function approach. Specially, there are two main findings from our calculation: (1) the thermal conductivity can reach a high value of 310 Wm$^{-1}$K$^{-1}$ in a 100 nm polyethylene chain at room temperature and the thermal conductivity increases with the length of the chain; (2) the Young's modulus in the polyethylene chain is as high as 374.5 GPa, and the polyethylene chain can sustain $32.85\% \pm 0.05\%$ (ultimate) strain before undergoing structural phase transition into gaseous ethylene.
\end{abstract}

\pacs{66.30.hk, 66.25.+g, 63.22.-m, 62.20.de, 73.22.-f}
\keywords{polyethylene chain, thermal conductivity, phonon vibration, ultimate strain, stiffness}
\maketitle

\pagebreak

Polymer is a useful organic material in many important fields such as solar cell,\cite{MillerNC} where high thermal conductivity will be helpful to deliver heat promptly and prevent the device from being broken down by localized heating. However, it is well known that the polymer is a poor thermal conductor with thermal conductivity less than 1 Wm$^{-1}$K$^{-1}$ due to a huge number of interfaces inside the material.\cite{DavidDJ} In recent years, there have been increasing efforts to enhance the thermal transport probability of polymer. A straightforward method is to disperse some materials of high thermal conductivity (such as carbon nanotube or graphene) into polymer, which forms the polymer composite. However, at small loading fractions of graphene or carbon nanotubes, the overall thermal conductivity of the polymer remains relatively low because the heat transport is still strongly affected by the interfaces in the composite.\cite{GoyalV,ShahilKMF} Recently, Liu and Yang found that the thermal conductivity of the polymer can be considerably increased by mechanical strain.\cite{LiuJ} The enhancement of the thermal conductivity is attributed to a better alignment of polymer chains under mechanical strain. It implies that a more efficient way to enhance the thermal conductivity is to produce a crystalline polymer, as confirmed by a recent study.\cite{NiB} The properties of the crystalline polymer are closely related to the single polymer chain, and thus it is meaningful to investigate a single polymer chain. Furthermore, it may be possible to prepare a single polymer chain in the laboratory, considering the rapid development of the experimental detection technique.\cite{AltoeV} As polyethylene is the most widely applied polymer, it is of practical significance to investigate the properties of a single polyethylene chain. Henry and Chen applied the molecular dynamics simulation to study the thermal transport in the polyethylene chain.\cite{HenryA2008,HenryA2009} We are more interested in an accurate prediction for the upper limit of the thermal conductivity in the polyethylene chain, since it is almost an impossible task to mimic precisely the complex experimental set up in a theoretical theme.

\begin{figure}[htpb]
  \begin{center}
    \scalebox{1.0}[1.0]{\includegraphics[width=8cm]{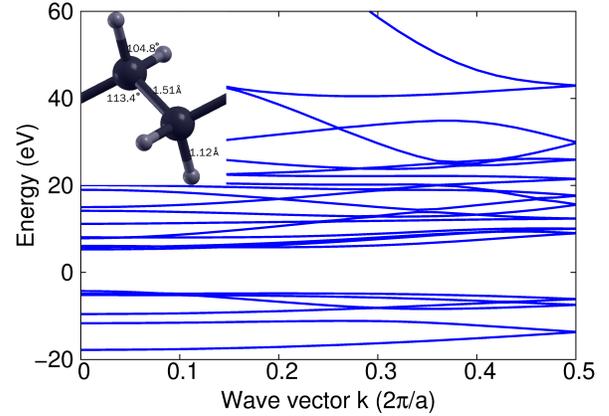}}
  \end{center}
  \caption{(Color online) Electronic band structure for the polyethylene chain. The energy is with reference to the Fermi level. Inset shows the relaxed configuration of a unit cell ${\rm C_{2}H_{4}}$.}
  \label{fig_electron}
\end{figure}
In this paper, we study the thermal transport and mechanical properties of the polyethylene chain by applying the {\it ab initio} calculation and the non-equilibrium Green's function (NEGF) approach. The obtained thermal conductivity serves as an upper limit of the experimentally measured thermal conductivity for this system. For a polyethylene chain of 100 nm, the predicted thermal conductivity is as high as 310 Wm$^{-1}$K$^{-1}$ at room temperature. Our calculation shows that a single polyethylene chain is of high strength with Young's modulus $Y=374.5$ GPa and ultimate strain $\epsilon_{c}=32.85\% \pm 0.05\%$.

\begin{figure}[htpb]
  \begin{center}
    \scalebox{1.0}[1.0]{\includegraphics[width=8cm]{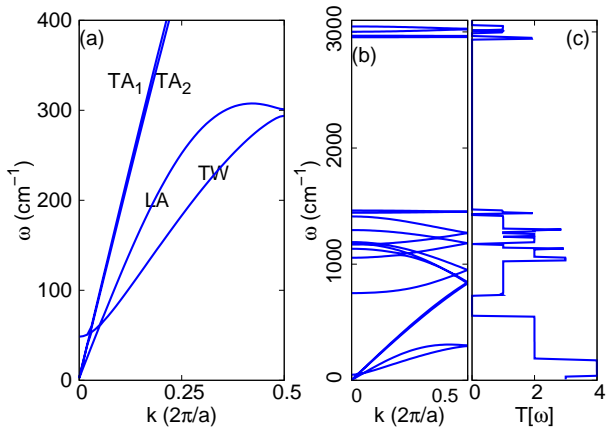}}
  \end{center}
  \caption{(Color online) Phonon dispersion and Young's modulus of polyethylene. (a). Four low-frequency branches of the phonon dispersion. (b). The phonon dispersion in whole frequency range. (c). The transmission from NEGF approach.}
  \label{fig_phonon}
\end{figure}
In the {\it ab initio} calculation, we use the SIESTA package\cite{SolerJM} to optimize the structure of the polyethylene chain. The local density approximation is applied to account for the exchange-correlation function with Ceperley-Alder parametrization\cite{CeperleyDM} and the double-$\zeta$ basis set orbital (DZ) is adopted. During the conjugate-gradient optimization, the maximum force on each atom is smaller than 0.005 eV/\AA. A mesh cut off of 120 Ry is used. Periodic boundary condition is applied in the growth direction, while free boundary conditions are applied in the two transverse directions by introducing sufficient vacuum space. There are two types of calculations in this work. In the calculation of the electron and phonon band structures, a unit cell ${\rm C_{2}H_{4}}$ is used and an $8\times 1\times 1$ Monkhorst-Pack k-grid is chosen for the sampling of the one-dimensional Brillouin zone. In the investigation of the Young's modulus and the thermal transport, a finite polyethylene chain of 20 unit cells is calculated and the Gamma point $k$ sampling is adopted.

\begin{figure}[htpb]
  \begin{center}
    \scalebox{0.8}[0.8]{\includegraphics[width=8cm]{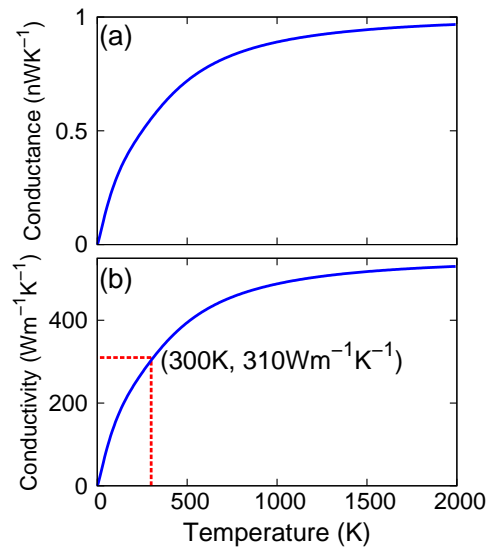}}
  \end{center}
  \caption{(Color online) Thermal transport properties in the polyethylene chain. (a). Thermal conductance versus temperature. (b). Thermal conductivity ($\kappa$) calculated from thermal conductance ($\sigma$) by: $\kappa=\sigma L/s$, with cross-sectional area $s=18.24$~{\AA}. The length $L=100$~nm.}
  \label{fig_conductance}
\end{figure}
\begin{figure*}[htpb]
  \begin{center}
    \scalebox{0.8}[0.8]{\includegraphics[width=\textwidth]{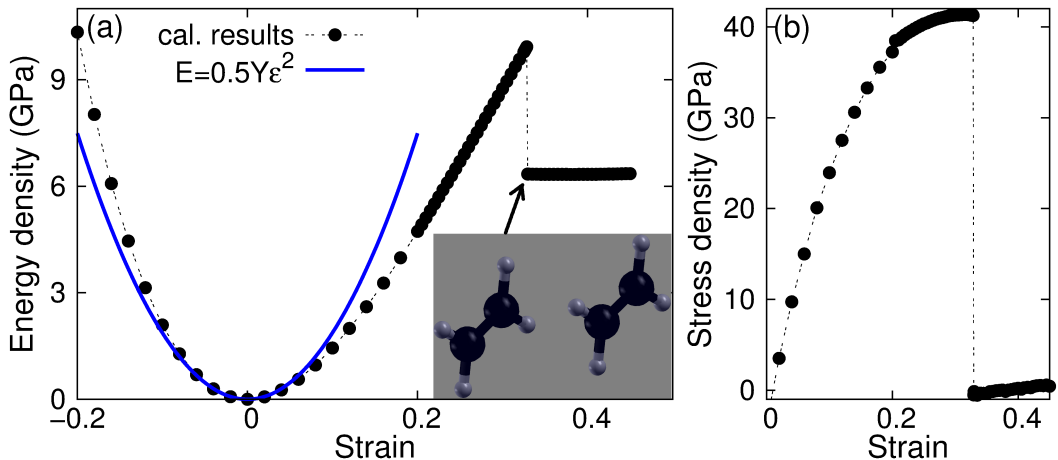}}
  \end{center}
  \caption{(Color online) Mechanical properties of the polyethylene chain. (a). The density of strain energy versuses strain. The Young's modulus $Y=374.5$ Gpa is obtained by fitting strain energy density to $E=0.5Y\epsilon^{2}$ in small strain region $\epsilon\in[-0.01, 0.01]$. A big jump in the curve reveals the ultimate strain to be $\epsilon_{c}=32.85\% \pm 0.05\%$. Inset shows two neighboring ethylene moleculars after the phase transition of the polyethylene chain at $\epsilon=\epsilon_{c}$. (b). Stress-strain relation.}
  \label{fig_young}
\end{figure*}
The inset of Fig.~\ref{fig_electron} shows the optimized configuration of a ${\rm C_{2}H_{4}}$ unit cell in the polyethylene chain. The C-C bond length is 1.51~{\AA}, while the C-H bond length is 1.12~{\AA}. The two angles are $\theta_{\rm CCC}=113.4^{\circ}$ and $\theta_{\rm HCH}=104.8^{\circ}$. These structure parameters are in good agreement with the experimental results for the crystalline polyethylene.\cite{ShearerHMM,TearePW,MathisenH,KaveshS} Fig.~\ref{fig_electron} is the electronic band structure of the polyethylene chain, yielding a large band gap of 9.58 eV, which coincides with the experimental value of 8.8 eV\cite{LessKJ} and other {\it ab initio} calculation\cite{MiaoMS}. The large band gap in the polyethylene chain is related to the tetrahedral (sp$^{3}$) structure of the carbon atom in this material, inducing a strong binding for all electrons. Fig.~\ref{fig_phonon}~(a) displays the phonon dispersion in the low-frequency range. The wave vector $k$ is along the growth direction and in the unit of $2\pi/a$, with $a=2.53$~{\AA} as the size of the unit cell ${\rm C_{2}H_{4}}$. The twisting (TW) phonon has nonzero frequency at $k=0$. This is due to the loss of the rigid rotational invariance symmetry\cite{JiangJW2006} in the {\it ab initio} calculation. For the three translational acoustic phonons, the sound velocities are 9.35 km/s for the longitudinal acoustic (LA) phonon, and 14.14 km/s and 15.03 km/s for the two transverse acoustic (TA) phonons. It is quite interesting that the sound velocities for TA phonons are faster than those for the LA phonon. This phenomenon implies that the polyethylene chain is more stiff in the transverse directions, owing to its zigzag configuration formed by C-C bonds along the growth direction. It should be noted that these three sound velocities are actually quite large and very close to that of the single-walled carbon nanotube, where the LA and TA sound velocities are around 20 km/s and 10 km/s, respectively.\cite{JiangJW2008} The full phonon dispersion is demonstrated in Fig.~\ref{fig_phonon}~(b) for the polyethylene chain. There are four typical regions. Below 400 cm$^{-1}$ are the acoustic phonons, as zoomed in in panel (a). The frequency of the C-C vibration is around 1300 cm$^{-1}$, which is quite close to that in diamond,\cite{BohningMS} due to a similar sp$^{3}$ bonding in both systems. The C-H bond is characterized by its vibration frequency around 1500 cm$^{-1}$. The highest frequency around 3000 cm$^{-1}$ corresponds to the vibration of the H atom, which is also observed in the Raman measurement (see eg. Ref.~\onlinecite{StuartBH}).

Fast sound speeds indicate strong thermal transport capability of the polyethylene chain. The thermal transport can be investigated by different methods. Classical results can be obtained from the molecular dynamics simulation of the thermal transport, where the phonon-phonon scattering dominates the transport property.\cite{HenryA2008,HenryA2009} From the theoretical point of view, it is a big challenge to provide accurate prediction for the thermal conductivity, because the samples in the experiment always possess various unpreventable defects. Hence, a more practical task is to provide an accurate (quantum) prediction for the upper limit of the thermal conductivity. For this purpose, we can apply the ballistic NEGF method.\cite{WangJS2008,JiangJWjap} It is based on quantum mechanics. The phonon-phonon scattering is ignored in the ballistic transport region, which is actually quite reasonable for nano-materials. The combination of the NEGF and the {\it ab initio} calculation can provide us an accurate upper limit for the thermal conductivity.\cite{JiangJW2011} In the NEGF approach, the thermal conductance is calculated by the Landauer formula:
\begin{eqnarray}
\sigma & = & \frac{1}{2\pi}\int d\omega\hbar\omega T[\omega]\left[\frac{\partial n(\omega,T)}{\partial T}\right],
\end{eqnarray}
where $\hbar$ is the Planck's constant. $n(\omega,T)$ is the Bose-Einstein distribution function. The transmission $T[\omega]$ is obtained from the Caroli formula:
\begin{eqnarray}
T[\omega] & = & {\rm Tr}\left(G^{r}\Gamma_{L}G^{a}\Gamma_{R}\right),
\end{eqnarray}
where $G^{r}$ is the retarded Green's function. $G^{a}=\left(G^{r}\right)^{\dagger}$ is the advanced Green's function and $\Gamma_{L/R}$ is the self-energy. These Green's functions can be calculated from the force constant matrix from the {\it ab initio} calculation.\cite{JiangJW2011} Fig.~\ref{fig_phonon}~(c) shows the transmission function for the polyethylene chain. The function exhibits some regular steps, due to the absence of phonon-phonon scattering. Fig.~\ref{fig_conductance}~(a) shows the temperature dependence for the thermal conductance. The thermal conductance ($\sigma$) does not depend on the length of the system. It can be used to get the thermal conductivity ($\kappa$) of a polyethylene chain with arbitrary length $L$: $\kappa=\sigma L/s$, where $s$ is the cross-sectional area of the polyethylene chain. Fig.~\ref{fig_conductance}~(b) gives the thermal conductivity for a polyethylene chain of 100~nm. Particularly, the room-temperature thermal conductivity is as high as 310 Wm$^{-1}$K$^{-1}$. The thermal conductivity will increase with increasing length of the polyethylene chain, which is quite similar as the thermal conductivity of two-dimensional graphene.\cite{Balandin,NikaPRB,NikaAPL,Balandin2011,JiangJW2009} We should stress that this value serves as an upper limit for the thermal conductivity of the 100~nm polyethylene chain. If the experimental samples are of high quality, then the measured thermal conductivity should approach 310 Wm$^{-1}$K$^{-1}$ from lower side.

We note that the phonon-phonon scattering is ignored in the ballistic transport region, so the derived value of thermal conductivity only serves as an upper limit for the experimental value. In the ballistic transport region, thermal conductivity always increases linearly with increasing length without transition behavior, which has been discussed for graphene.\cite{Nikapssb,GhoshS} The thermal conductivity is over 3000 Wm$^{-1}$K$^{-1}$ for polyethylene chain of 1.0 micrometer long. This is a high but reasonable value, which is very close to the thermal conductivity of graphene of similar size.\cite{Balandin} This agreement implies that the phonon transport in the polyethylene chain can be regarded as ballistic transport up to very long length.

A fast sound speed in the polyethylene chain also implies high stiffness of this structure. Indeed, we find large value for the axial Young's modulus from Fig.~\ref{fig_young}~(a). In the figure, the density of the strain energy is calculated for a finite polyethylene chain of 20 unit cells under axial uniform strain. In the calculation of the volume, the cross-sectional area $s=18.24$~{\AA}$^{2}$ is chosen as half of the experimental value for the unit cell of crystalline polyethylene with two polyethylene chains.\cite{BunnCW} In the small strain region $\epsilon\in [-0.01, 0.01]$, the nonlinear effect is neglectable, so the elastic parameter can be obtained by fitting the strain energy to $E(\epsilon)=1/2Y\epsilon^2$ in this strain region. The obtained Young's modulus $Y=374.5$~GPa. This value is even higher than the Young's modulus of crystalline polyethylene around 300 GPa.\cite{HollidayL,ZwitnewbergA,MatsuoM,ZhaoJ} In larger strain region $\epsilon\in [-0.2, 0.2]$, the energy curve is obvious asymmetry under tensile and compressive strain, which illustrates a strong nonlinear effect for the polyethylene chain in the large strain region. This nonlinear effect holds such property that the interaction is stronger for compressive strain than that of the tensile strain, which will result in a positive thermal expansion phenomenon during heating. With further increase of the tensile strain, the strain energy density increases linearly and the polyethylene chain is broken at a critical strain $\epsilon_{c}=32.85\% \pm 0.05\%$, which serves as the ultimate strain of the polyethylene chain. The jump of the strain energy density at $\epsilon=\epsilon_{c}$ corresponds to a bond energy of 1.031 eV for each sp$^{3}$ C-C bond in the polyethylene chain. The discontinuity of the energy jump is a characteristic feature for the broken of a single chain. In the structural phase transition, the sp$^{3}$ bonds in the polyethylene are broken and new sp2 bonds are formed. According to this bond transition, the polyethylene chain changes into 20 gaseous ethylene moleculars. Two ethylene moleculars are shown in the inset of Fig.~\ref{fig_young}~(a) after the phase transition of the polyethylene chain. The C-C and C-H bond lengths are 1.332~{\AA} and 1.103~{\AA}, respectively. The angles are $\theta_{HCC}=122.2^{\circ}$ and $\theta_{HCH}=115.7^{\circ}$. These structure parameters agree quite well with the experimental data.\cite{BartellLS} The distance between two neighboring ethylene moleculars is about 3.356~{\AA}, which falls in a typical interaction range of the van der Waals potential. The phase transition phenomenon can also be well captured by the stress-strain relation shown in Fig.~\ref{fig_young}~(b), where the stress reaches a saturate value before $\epsilon_{c}$ and suddenly jumps to a very small value at $\epsilon=\epsilon_{c}$.

In conclusion, we have performed {\it ab initio} calculation to study the phonon, mechanical, and thermal transport properties of the polyethylene chain. Our results show that the sound speeds are very high in this system: 9.35 km/s for the LA phonon, and 14.14 km/s and 15.03 km/s for the two transverse phonons. Using the NEGF quantum approach, we provide an upper limit for the thermal conductivity of the polyethylene chain of any length; particularly, the room-temperature thermal conductivity can be as high as 310 Wm$^{-1}$K$^{-1}$ for a 100 nm polyethylene chain. In consistent with the high speed, we get a large Young's modulus (374.5 GPa), which is even larger than that of the crystalline polyethylene. After increasing the applied tensile strain, we observe the structural transition from a polyethylene chain into gaseous ethylene, yielding an ultimate strain as high as $\epsilon_{c}=32.85\% \pm 0.05\%$ for the polyethylene chain.

\textbf{Acknowledgements} The work is supported by the Grant Research Foundation (DFG) Germany and by Academic Research Fund Tier 1 from Ministry of Education, Singapore (Grant No. 401050000).

\end{document}